\documentclass[a4paper,fleqn,usenatbib]{mnras}

\usepackage{txfonts}
\usepackage{threeparttable}

\usepackage[T1]{fontenc}
\usepackage{ae,aecompl}
\usepackage{comment}

\usepackage{graphicx}   
\usepackage{mathabx}    

\title[A dense K2 hot-Jupiter]{K2-113b: A dense hot-Jupiter transiting a solar analogue}

\author[Espinoza et al.]
{N\'estor Espinoza$^{1,2}$, 
Markus Rabus$^1$, 
Rafael Brahm$^{1,2}$, 
Mat\'ias Jones$^3$, 
Andr\'es Jord\'an$^{1,2,4}$,
\newauthor
Felipe Rojas$^1$,
Holger Drass$^1$, 
Maja Vu\v{c}kovi\'{c}$^5$, 
Joel D. Hartman$^6$, 
James S. Jenkins$^7$,
\newauthor
Cristi\'an Cort\'es$^{8,2}$
\\
$^{1}$ Instituto de Astrof\'isica, Facultad de F\'isica, Pontificia Universidad Cat\'olica de Chile,\\
Av. Vicu\~na Mackenna 4860, 782-0436 Macul, Santiago, Chile.\\
$^{2}$ Millennium Institute of Astrophysics, Av. Vicu\~na Mackenna 4860, 782-0436 Macul, Santiago, Chile.\\
$^{3}$ European Southern Observatory, Alonso de Cordova 3107, Vitacura, Casilla 19001, Santiago 19, Chile.\\
$^{4}$ Max-Planck-Institut f\"ur Astronomie, K\"onigstuhl 17, Heidelberg 69 117, Germany.\\
$^{5}$ Instituto de F\'isica y Astronom\'ia, Facultad de Ciencias, Universidad de Valpara\'iso, Gran Breta\~na 1111, \\ 
Playa Ancha, Valpara\'iso 2360102, Chile.\\
$^{6}$ Department of Astrophysical Sciences, Princeton University, NJ 08544, USA.\\
$^{7}$ Departamento de Astronom\'ia, Universidad de Chile, Camino al Observatorio 1515, Cerro Cal\'an, Santiago, Chile.\\
$^{8}$ Departamento de F\'isica, Facultad de Ciencias B\'asicas, Universidad Metropolitana de la Educaci\'on, Av. Jos\'e \\
Pedro Alessandri 774, 7760197, \~Nu\~noa, Santiago, Chile.
}

\date{Accepted XXX. Received YYY; in original form ZZZ}

\pubyear{2016}

\begin{document}
\label{firstpage}
\pagerange{\pageref{firstpage}--\pageref{lastpage}}
\maketitle

\begin{abstract}
We present the discovery of K2-113b, a dense hot-Jupiter discovered using photometry from Campaign 8 of the Kepler-2 (K2) mission and 
high-resolution spectroscopic follow up obtained with the FEROS spectrograph. The planet orbits a $V=13.68$ solar analogue in a $P=5.81760^{+0.00003}_{-0.00003}$ 
day orbit, has a radius of $0.93^{+0.10}_{-0.07}R_J$ and a mass of $1.29^{+0.13}_{-0.14}M_J$. With a density of $1.97^{+0.60}_{-0.53}$ gr/cm$^3$, 
the planet is among the densest systems known having masses below 2 $M_J$ and $T_\textnormal{eq} > 1000$, and is just above the temperature limit at which 
inflation mechanisms are believed to start being important. Based on its mass and radius, we estimate that K2-113b should have a heavy element 
content on the order of $\sim$ 110 $M_{\oplus}$ or greater.
\end{abstract}




\section{Introduction}
Transiting extrasolar planets are one of the most precious systems to discover because they allow for a wide range of characterization possibilities. 
Combined with radial velocity or transit timing variation analysis, the mass of these systems can be extracted, which in turn allow us to 
compute their densities, an important measurement that sheds light on the composition of these distant worlds. 

Despite their importance, only a small fraction ($\sim 10\%$) of the currenlty $\sim 2500$ known transiting extrasolar 
planets\footnote{\url{http://www.exoplanets.org}, retrieved on 2016/11/19.} are well suited for further characterization studies, mainly because the 
bulk of these discoveries have been made with the original \textit{Kepler} mission \citep{borucki:2010}, whose stars are generally too faint and most 
of the planets too small to characterize. {Although the bulk of the transiting extrasolar planets fully characterized to date 
come from ground-based transit surveys such as HATNet \citep{HATNet}, HATSouth \citep{HATSouth} and WASP \citep{WASP}}, the search for transiting 
exoplanets around relatively bright stars {has also benefited from the discoveries made by the} repurposed \textit{Kepler} mission, 
dubbed K2, {which has allowed to push discoveries even to smaller planets}, with hundreds of new systems discovered to 
date\footnote{\url{keplerscience.arc.nasa.gov}} \citep[see, e.g., ][and references therein]{crossfield:2016} and many more to come.

Among the different types of transiting extrasolar planets known to date, short-period ($P\lesssim10$), Jupiter-sized exoplanets -- 
the so-called ``hot-Jupiters" -- have been one of the most studied, mainly because they are the easiest to detect and characterize. 
However, these are also one of the most intriguing systems to date. One of the most interesting properties of these planets is their 
``inflation", i.e., the fact that most of them are larger than what is expected from structure and evolution models of highly irradiated 
planets \citep{baraffe:2003,fortney:2007}. Although the inflation mechanism is as of today not well understood, at irradiation levels of 
about $2\times 10^8$ ergs/cm$^2$/s ($\sim 1000$ K) evidence suggests it stops being important \citep{kovacs:2010,miller:2011,demory:2011}. 
Planets cooler than this threshold, {which here we refer to as} ``warm" Jupiters, appear on the other hand more compact than 
pure H/He spheres, which in turn implies an enrichment in heavy elements that most likely makes them deviate from the composition of their host stars 
\citep{thorngren:2016}. 

Here we present a new planetary system which is in the ``hot" Jupiter regime, but whose structure resembles more that of a  ``warm" Jupiter: 
K2-113b, a planet $\sim 10\%$ smaller than Jupiter but $\sim 30\%$ more massive orbiting a star very similar to our Sun. The 
paper is structured as follows. In Section \ref{sec:data} we present the data, which includes photometry from Campaign 8 of the K2 mission and spectroscopic 
follow-up using the FEROS spectrograph. In Section \ref{sec:analysis} we present the analysis of the data. Section \ref{sec:discussion} present 
a discussion and Section \ref{sec:conclusions} our conclusions.


\section{Data}
\label{sec:data}
\subsection{K2 Photometry}
\begin{figure*}
   \includegraphics[scale=0.8]{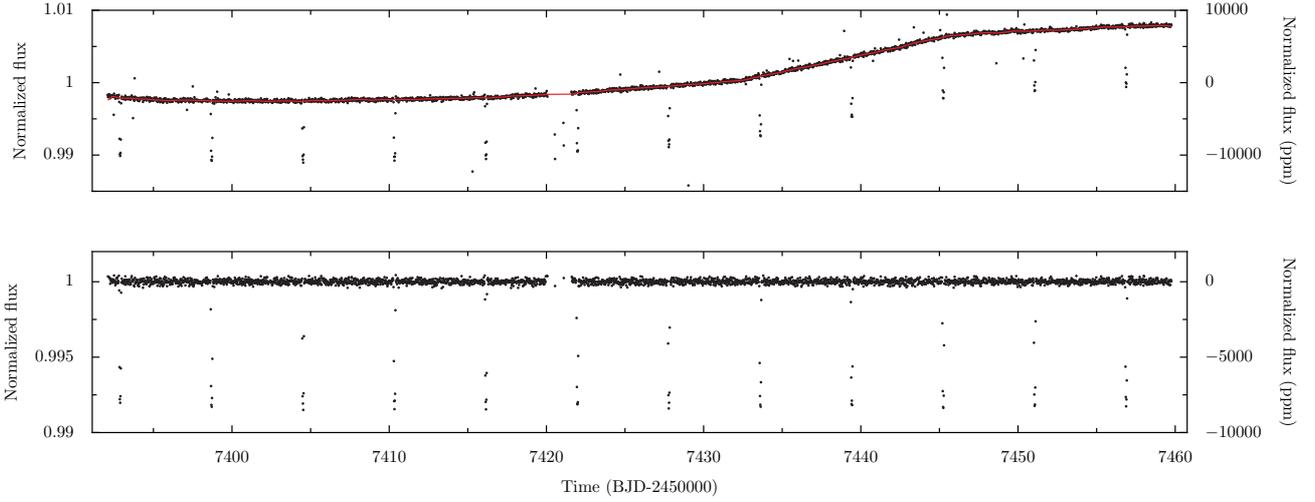}
    \caption{\textit{Upper panel}. EVEREST photometry for the star K2-113 (black points) along with the 20-hour median filter smoothed with a 
3-hour gaussian filter (red line), which captures the intrinsic variability of the star. \textit{Lower panel}. Photometry normalized with respect to 
our filter.}
    \label{fig:original_photometry}
\end{figure*}

The candidate selection for the photometry of Campaign 8 of the K2 mission was done as described in \cite{espinoza:2016}. Briefly,
the photometry is first normalized with respect to any long-term variation (either of instrumental and/or stellar nature) and 
candidates are selected using a Box Least Squares algorithm \citep[BLS;][]{kovacs:2002}. Here we decided to obtain the photometry 
for the candidate selection using our own implementation of the EVEREST algorithm described in \cite{luger:2016}, due to its potential 
of conserving stellar variability (which we filter for our candidate selection with a 20-hour median filter smoothed with a 3-hour gaussian filter, 
but which we also use in our analysis: see Section \ref{analysis}), although the full, final analysis performed here is done on the EVEREST lightcurve 
released at the MAST website\footnote{\url{https://archive.stsci.edu/prepds/everest/}}{, using the new updated method described 
in \cite{luger:2017}}. Our candidate selection procedure identified a 
planetary companion candidate to the star K2-113 (EPIC 220504338), with a period of $\sim$5.8 days and a depth of $\sim 7500$ ppm. 
The overall precision of the lightcurve is {$\sim 138$ ppm}; the photometry is shown in Figure \ref{fig:original_photometry}.

\subsection{Spectroscopic follow-up}
In order to confirm the planetary nature of our candidate, high-resolution spectroscopic follow-up was performed with the 
FEROS spectrograph \citep{feros:1998} mounted on the MPG 2.2m telescope located at La Silla Observatory in August (3 spectra) 
and November (6 spectra) of 2016, in order to obtain both initial stellar parameters for the candidate stellar host and 
high-precision radial velocity (RV) measurements. The spectra were obtained with the simultaneous calibration method, in which a ThAr calibration lamp is observed in a 
comparison fiber next to the science fiber, allowing us to trace instrumental RV drifts. The data was reduced with a dedicated pipeline 
\citep[CERES; ][]{jordan:2014,ceres:2016} which, in addition to the radial-velocities and bisector spans, also calculates rough atmospheric parameters 
for the target star. This indicated the candidate host star was a G dwarf, with an effective temperature of $T_\textnormal{eff}=5500 \pm 100$ K, 
surface log-gravity of $\log g_* = 4.2 \pm 0.3$ dex and a metallicity of $\textnormal{[Fe/H]}=0.2 \pm 0.1$ dex, all very much consistent with solar values.

\begin{figure}
   \includegraphics[width=\columnwidth]{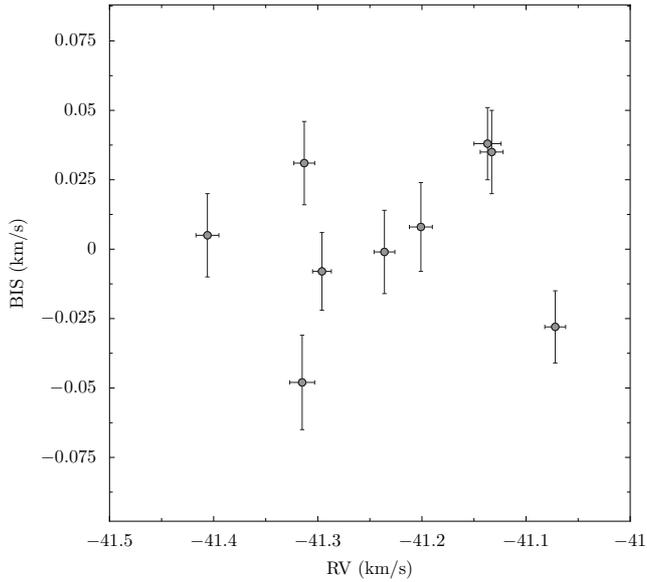}
    \caption{Radial velocity measurements versus the measured bisector spans in km/s obtained for our target star using the FEROS spectrograph. 
The correlation coefficient between these measurements is consistent with zero.}
    \label{fig:rvs_bis}
\end{figure}

The obtained RVs phased up nicely with the photometric ephemerides, hinting at a semi-amplitude of $\sim 140$ m/s, consistent with an object 
of planetary nature (see Section \ref{analysis}). In addition, the measured bisector spans (BIS) showed no correlation 
with the RVs, which is illustrated on Figure \ref{fig:rvs_bis}; performing a monte-carlo simulation by assuming the errors 
on the RVs and BIS are gaussian gives a correlation coefficient of $\rho = 0.15 \pm 0.16$, which is consistent with zero. The 
obtained radial velocities and bisector spans are presented in Table \ref{tab:rvs}. These results prompted us to perform a full 
analysis of the system, which we present in the next section.

\begin{table}
    \centering
    \caption{Radial velocities obtained with the FEROS spectrograph along with the measured bisector spans.}
    \label{tab:rvs}
    \begin{tabular}{lcccc} 
        \hline
        \hline
Time (BJD UTC)  & RV (km/s) & $\sigma_{\textnormal{RV}}$ & BIS (km/s) & $\sigma_{\textnormal{BIS}}$\\
        \hline
2457643.6804163 & -41.315 & 0.012 & -0.048 & 0.017 \\
2457645.7862304 & -41.236 & 0.010 & -0.001 & 0.015 \\
2457647.8755248 & -41.133 & 0.011 & 0.035 & 0.015 \\
2457700.7309840 & -41.201 & 0.011 & 0.008 & 0.016 \\
2457701.6407003 & -41.313 & 0.010 & 0.031 & 0.015 \\
2457702.7358066 & -41.406 & 0.011 & 0.005 & 0.015 \\
2457703.6386414 & -41.296 & 0.009 & -0.008 & 0.014 \\
2457704.6259996 & -41.137 & 0.013 & 0.038 & 0.013 \\
2457705.6178241 & -41.072 & 0.010 & -0.028 & 0.013 \\
        \hline
    \end{tabular}
\end{table}

\section{Analysis}
\label{sec:analysis}
\label{analysis}
\subsection{Stellar properties}
In order to obtain the parameters of the host star, we first use the Zonal Atmospheric Stellar Parameters Estimator \citep[ZASPE; ][]{brahm-zaspe:2016,zaspe:2016} algorithm. 
In brief, ZASPE compares the observed spectrum against a grid of stellar spectra in the most sensitive 
spectral zones to atmospheric parameters and determines the errors in the parameters by considering the systematic mismatch between the data and the models. 
In this case we run ZASPE on a high signal-to-noise (SNR; $\sim 100$) spectrum that was generated by co-adding the 9 individual FEROS spectra, which obtains 
a $T_\textnormal{eff} = 5627\pm 88$ K, $\log g_* = 4.400\pm 0.146$ dex, $[\textnormal{Fe/H}]=0.180\pm 0.062$ dex and projected rotational velocity 
$v\sin(i) = 2.06\pm 0.77$ km/s, which make the host star a (slightly metal-rich) solar analogue. 

In order to derive the radius, mass, age, luminosity and distance to the star, we used the latest 
version of the \texttt{isochrones} package \citep{isochrones:2015}, which uses the derived atmospheric parameters along with photometric data in order to 
estimate them with evolutionary tracks. The photometric data for our star was obtained from different sources; these are presented in Table 
\ref{tab:stellar_props}. We used the MESA Isochrones \& Stellar Tracks \citep[MIST; ][]{dotter:2016,choi:2016} instead of the 
Darthmouth \citep{dotter:2008} isochrones and stellar tracks, as the former cover wider ranges of radius, mass and age (although 
both gave results which were consistent within the errors). In order to explore the parameter space, the MULTINEST \citep{multinest:2009} algorithm as 
implemented in PyMultinest \citep{buchner:2014} was used because it is well suited for problems like the one at hand, which are inherenty degenerate. 
The derived stellar properties are presented in Table \ref{tab:stellar_props}, all of which are consistent with the star being very similar to our own Sun 
($R_* = 1.047^{+0.11}_{-0.08}R_\Sun$, $M_* = 1.007^{+0.040}_{-0.039}$, $L_* = 1.02^{+0.24}_{-0.18}$). As can be observed, the only parameter that significatnly 
deviates (at 3-sigma) from that of a ``solar twin" is the metallicity which, as mentioned above, is slighlty super-solar. We therefore consider the star a 
solar analogue.

\begin{table}
 \centering 
 \begin{center}
\caption{Stellar parameters of K2-113.}
 \label{tab:stellar_props}
 \begin{threeparttable}
  \centering
  \begin{tabular}{ lcr }
   \hline
   \hline
     Parameter &  Value & Source \\
   \hline
Identifying Information\\
~~~EPIC ID & 220504338 & EPIC\\
~~~2MASS ID & 01174783+0652080 & 2MASS\\
~~~R.A. (J2000, h:m:s) & 01$^h$17$^m$47.829$s$ & EPIC\\
~~~DEC (J2000, d:m:s) & +06$^o$52$'$08.02$''$ & EPIC\\
~~~R.A. p.m. (mas/yr)  & $22.3\pm1.7$ & UCAC4\\
~~~DEC p.m. (mas/yr) & $-15.8\pm4.1$ & UCAC4\\
Spectroscopic properties\\
~~~$T_\textnormal{eff}$ (K) & $5627\pm 88$ & ZASPE\\
~~~Spectral Type & G & ZASPE\\
~~~[Fe/H] (dex) & $0.180\pm 0.062$ & ZASPE\\
~~~$\log g_*$ (cgs)& $4.400\pm 0.146$ & ZASPE\\
~~~$v\sin(i)$ (km/s)& $2.06\pm 0.77$ & ZASPE\\
Photometric properties\\\
~~~$K_p$ (mag)& 13.51 & EPIC\\
~~~$B$ (mag)& $14.445\pm 0.050$ & APASS\\
~~~$V$ (mag)& $13.684\pm 0.030$ & APASS\\
~~~$g'$ (mag)& $14.016\pm0.030$ & APASS\\
~~~$r'$ (mag)& $13.459\pm0.020$ & APASS\\
~~~$i'$ (mag)& $13.299\pm0.030$ & APASS\\
~~~$J$ (mag)& $12.347\pm0.023$ & 2MASS\\
~~~$H$ (mag)& $11.998\pm0.025$ & 2MASS\\
~~~$Ks$ (mag)& $11.949\pm0.021$ & 2MASS\\
Derived properties\\
\vspace{0.1cm}
~~~$M_*$ ($M_\Sun$)& $1.007^{+0.040}_{-0.039}$ & \texttt{isochrones}*\\
\vspace{0.1cm}
~~~$R_*$ ($R_\Sun$)& $1.047^{+0.11}_{-0.08}$ & \texttt{isochrones}*\\
\vspace{0.1cm}
~~~$\rho_*$ (g/cm$^3$)& $1.23^{+0.36}_{-0.32}$ & \texttt{isochrones}*\\
\vspace{0.1cm}
~~~$L_*$ ($L_\Sun$)& $1.02^{+0.24}_{-0.18}$ & \texttt{isochrones}*\\
\vspace{0.1cm}
~~~Distance (pc)& $553.4^{+59.0}_{-43.0}$ & \texttt{isochrones}*\\
\vspace{0.1cm}
~~~Age (Gyr)& $5.9^{+2.7}_{-3.4}$ & \texttt{isochrones}*\\
   \hline

   \end{tabular}
      \textit{Note}. Logarithms given in base 10.\\
      *: Using stellar parameters obtained from ZASPE.
  \end{threeparttable}
 \end{center}
 \end{table}

\subsection{Planet scenario validation}

We performed a blend analysis following {\cite{hartman:2011,hartman:2011a}}, which attempts to model the available light curves, photometry calibrated to an absolute 
scale, and spectroscopically determined stellar atmospheric parameters, using combinations of stars with parameters constrained to lie on the 
\cite{girardi:2000} evolutionary tracks. Possible blend scenarios include blended eclipsing binary (bEB) and hierchical 
triple (hEBs) systems. The analysis includes fits of the secondary eclipses and out of transit variations using the photometric data. We find that 
the data are best described by a planet transiting a star. All of the above mentioned blend scenarios are rejected at more than
$5-\sigma$ using the photometry alone. Including the RV data, the scenarios are further ruled out: the simulated RVs for the blend model that 
provides the best fit to the photometric data imply variations in RV on the order of 500 m/s, which are much higher than what we observe. 
Based on this analysis, we consider our planet validated. 

It is important to note that with our validation procedures, we can't rule out the possibility that the planetary transit is being diluted by a 
star whithin the 12'' {aperture radius used to obtain the K2 photometry}. However, there is no known blending source within this 
radius in catalogs such as the Gaia Data Release 1 \citep{gaia:2016} and UCAC4 \citep{zacharias:2013}, while there is some stars within the 12'' 
aperture in the Sloan Digital Sky Survey (SDSS) that have $g>20$, which would produce negligible dilutions in the K2 lightcurve.

\subsection{Joint analysis}
As in \cite{brahm:2016}, the joint analysis of the K2 photometry and the FEROS RVs was performed using the EXOplanet traNsits and 
rAdIal veLocity fittER \citep[EXONAILER; ][]{espinoza:2016} algorithm, which is available at GitHub\footnote{\url{https://github.com/nespinoza/exonailer}}. 
The algorithm makes use of the \texttt{batman} package \citep{batman:2015} in order to perform the transit modelling, which has the advantage of allowing 
the usage of any limb-darkening law, which has been proven to be of importance if unbiased transit parameters are to be retrieved from high-precision 
photometry \citep{EJ15}. As recommended in the study of \cite{EJ15}, we decided to let the limb-darkening coefficients be free parameters in the fit. 
Following the procedures outlined in \cite{EJ16}, we concluded that the square-root law is the optimal law to use in our case{, as 
this is the law that retrieves the smaller mean-squared error (i.e., the best bias/variance trade-off) on the planet-to-star radius ratio, which is 
the most important parameter to derive for this exoplanet, as it defines the planetary density. The mean-squared error was estimated by sampling lightcurves 
with similar geometric, noise and sampling properties as the observed transit lightcurve, taking the spectroscopic information in order to model the real, 
underlying limb-darkening effect of the star. }We sampled the 
coefficients of this law in our joint analysis using the efficient uninformative sampling scheme outlined in \cite{kipping:2013}. {In 
order to take into account the smearing of the lightcurve due to the $\sim 27$ minute ``exposures" of the Kepler long-cadence observations, we use the 
selective resampling technique described in \cite{resampling} with $N=20$ resampled points per data-point in our analysis.} 

\begin{figure}
   \includegraphics[width=\columnwidth]{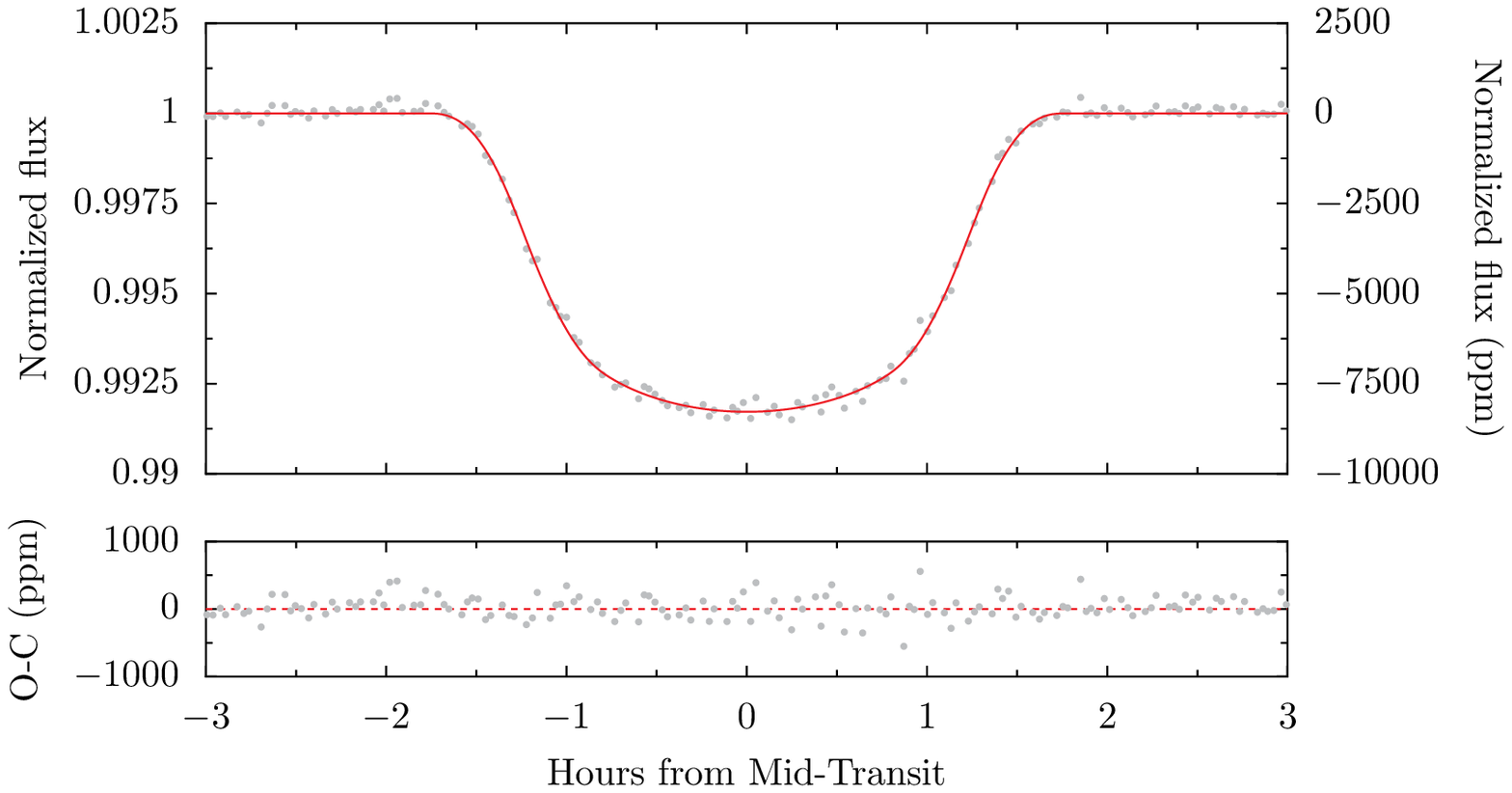}
    \caption{Phase-folded K2 photometry along with the best-fit transit model obtained from the joint analysis peformed with the EXONAILER algorithm.}
    \label{fig:transit}
\end{figure}

\begin{figure}
   \includegraphics[width=\columnwidth]{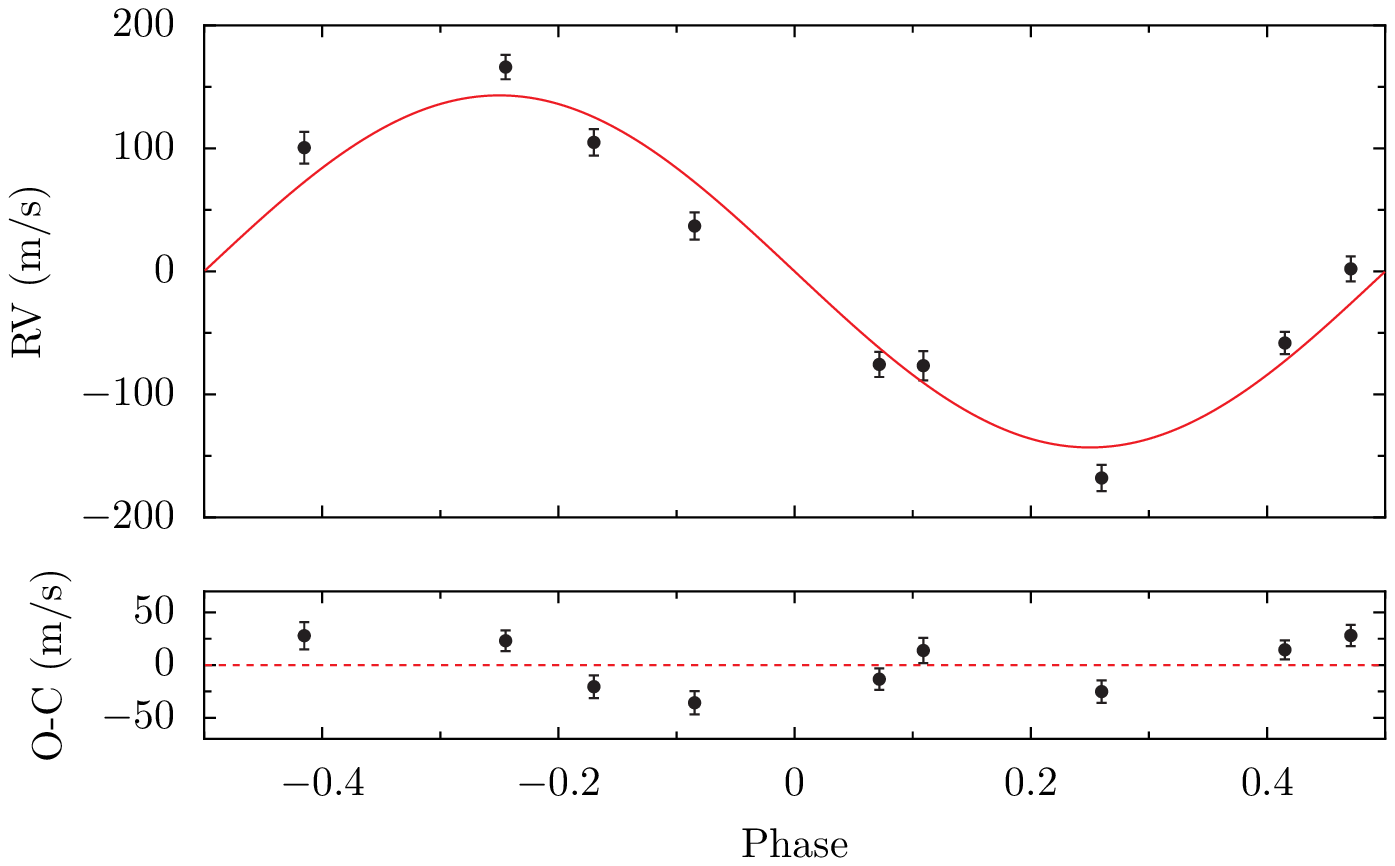}
    \caption{Phase-folded FEROS radial-velocities along with the best-fit model obtained from the joint analysis peformed with the EXONAILER algorithm.}
    \label{fig:rvs}
\end{figure}

{The RV analysis in our exonailer fit includes a radial-velocity jitter term that is added in quadrature to the measured uncertainties. We tried 
both, circular and non-circular models, but computing the BIC-based evidence of both models the non-circular model was indistinguishable from the circular 
one (the evidence for the non-circular model was $\sim 1.4$ that of the circular model). As such, we decided to use the most parsimonious model of both, 
and fixed the eccentricity to zero. We note that the non-circular model allowed us to put a 3-sigma upper limit on the eccentricity of $e<0.13$.} 
Figure \ref{fig:transit} shows the phase-folded photometry along with the best-fit model and Figure \ref{fig:rvs} shows the radial-velocity 
measurements and the corresponding best-fit model from our joint analysis. Table \ref{tab:planet-params} presents the retrieved parameters. 
Note the moderate jitter of the star, on the order of $\sigma_{\textnormal{RV}}\sim 20$ m/s. As can be observed, the planet has a radius of 
$R_p=0.93^{+0.10}_{-0.07}R_\textnormal{J}$, and a mass of $M_p=1.29^{+0.13}_{-0.14}M_\textnormal{J}$, giving a density of $1.97^{+0.60}_{-0.53}$ gr/cm$^3$ 
for this planet, wich is on the high side when compared to a ``typical" hot-Jupiter (where $\rho_p \lesssim 1$ g/cm$^3$). We discuss these 
planetary parameters in the context of the discovered exoplanets in the next section.

\begin{table*}
 \caption{Orbital and planetary parameters for K2-113b.}
 \label{tab:planet-params}
 \begin{threeparttable}
  \centering
  \begin{tabular}{ lcl }
   \hline
   \hline
     Parameter &  Prior & Posterior Value \\
   \hline
Lightcurve parameters\\
\vspace{0.1cm}
~~~$P$ (days)\dotfill    & $\mathcal{N}(5.8177,0.1)$ & 5.817608$^{+0.000031}_{-0.000029}$ \\
\vspace{0.1cm}
~~~$T_0-2450000$ (BJD)\dotfill    & $\mathcal{N}(7392.88575,0.1)$ & 7392.88605$^{+0.00019}_{-0.00019}$ \\
\vspace{0.1cm}
~~~$a/R_{\star}$ \dotfill    &$\mathcal{U}(1,15)$& $11.39^{+0.32}_{-0.32}$ \\
\vspace{0.1cm}
~~~$R_{p}/R_{\star}$\dotfill    & $\mathcal{U}(0.01,0.2)$ & $0.0911^{+0.00070}_{-0.00083}$ \\
\vspace{0.1cm}
~~~$i$ (deg)\dotfill & $\mathcal{U}(70,90)$ &86.21$^{+0.20}_{-0.21}$\\
\vspace{0.1cm}
~~~$q_1$ $^a$ \dotfill & $\mathcal{U}(0,1)$&$0.72^{+0.17}_{-0.16}$\\
\vspace{0.1cm}
~~~$q_2$ $^a$\dotfill & $\mathcal{U}(0,1)$& $0.33^{+0.11}_{-0.15}$\\
\vspace{0.1cm}
~~~$\sigma_w$ (ppm) \dotfill & $\mathcal{J}(10,500)$ & 138.4$^{+1.8}_{-1.6}$\\
\vspace{0.1cm}
RV parameters\\
\vspace{0.1cm}
~~~$K$ (m s$^{-1}$)\dotfill   & $\mathcal{N}(0,100)$ & $144.5^{+13.5}_{-15.4}$\\
\vspace{0.1cm}
~~~$\mu$ (km s$^{-1}$)\dotfill    & $\mathcal{N}(-41.24,0.01)$ & $-41.2375^{+0.0063}_{-0.0065}$ \\
\vspace{0.1cm}
~~~$\sigma_{\textnormal{RV}}$ (m s$^{-1}$)\dotfill    & $\mathcal{J}(1,100)$ & $24.3^{+10.1}_{-6.6}$ \\
\vspace{0.1cm}
~~~$e$ \dotfill    & --- & $0$ (fixed; {upper limit$^c$ $e<0.13$}) \\
\vspace{0.1cm}
Derived Parameters\\
\vspace{0.1cm}
~~~$s_1$ $^a$ \dotfill &  \dotfill & $0.30^{+0.20}_{-0.19}$\\
\vspace{0.1cm}
~~~$s_2$ $^a$\dotfill &   \dotfill & $0.55^{+0.26}_{-0.29}$\\
\vspace{0.1cm}
~~~$M_p$ ($M_\textnormal{J}$)       \dotfill      &---& $1.29^{+0.13}_{-0.14}$  \\
\vspace{0.1cm}
~~~$R_p$ ($R_\textnormal{J}$)       \dotfill       &---& $0.93^{+0.10}_{-0.07}$  \\
\vspace{0.1cm}
~~~$\rho_p$ (g/cm$^3$)       \dotfill      &---& $1.97^{+0.61}_{-0.53}$  \\
\vspace{0.1cm}
~~~$\log g_p$ (cgs)             \dotfill      &---& $3.56^{+0.08}_{-0.10}$  \\
\vspace{0.1cm}
~~~$a$ (AU)             \dotfill      &---& $0.0558^{+0.0059}_{-0.0049}$  \\
\vspace{0.1cm}
~~~$\log \langle F \rangle$ (cgs)$^b$      \dotfill      &---& $8.640^{+0.033}_{-0.034}$  \\
\vspace{0.1cm}
~~~$V_\textnormal{esc}$ (km/s)             \dotfill      &---& $69.7^{+4.9}_{-4.8}$  \\
\vspace{0.1cm}
~~~$T^d_\textnormal{eq}$ (K)  \dotfill          &&\\
\vspace{0.1cm}
~~~\ Bond albedo of $0.0$     &---& $1178^{+22}_{-23}$  \\
\vspace{0.1cm}
~~~\ Bond albedo of $0.75$        &---&     $833^{+16}_{-16}$  \\
   \hline
   \end{tabular}
   \textit{Note}. Logarithms given in base 10. $\mathcal{N}(\mu,\sigma)$ stands for a normal prior with mean $\mu$ and standard-deviation
   $\sigma$, $\mathcal{U}(a,b)$ stands for a uniform prior with limits $a$ and $b$ and $\mathcal{J}(a,b)$ stands for a Jeffrey's prior
   with the same limits. Times are given in BJD TDB.\\
   $^a${The $q_1$ and $q_2$ parameters are the} triangular sampling coefficients used to fit for the square-root limb-darkening 
   law \citep{kipping:2013}. The $s_1$ and $s_2$ limb-darkening coefficients {are} recovered by the transformation 
   $s_1 = \sqrt{q_1}(1-2q_1)$ and $s_2 = 2\sqrt{q_1}q_2$.\\
   $^b$Orbit averaged incident stellar flux on the planet.\\
   {$^c$3-sigma upper limit obtained from a non-circular joint fit to the data (see text).}\\
   {$^d$Full energy redistribution has been assumed.}\\
  \end{threeparttable}
 \end{table*}

\subsection{Searching for additional signals in the K2 photometry}
\begin{figure}
   \includegraphics[width=\columnwidth]{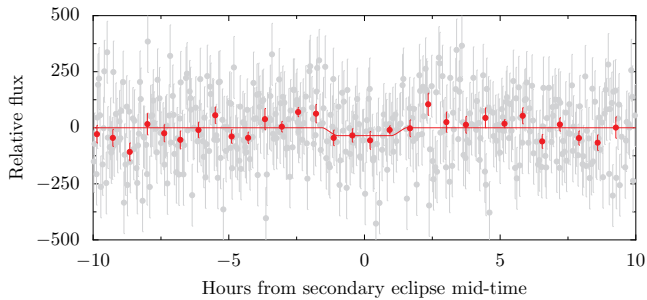}
    \caption{Secondary eclipse constrain using the K2 photometry (grey points; red points indicate binned data using 40-minute bins in phase-space, 
which have median errorbars of 35 ppm per point). The best-fit eclipse depth using this data (red solid line) is $F_p/F_* = 35.7^{+16.8}_{-19.0}$ 
ppm.}\label{fig:sec_plot}
\end{figure}

The K2 photometry was inspected in order to search for additional transiting planets, secondary eclipses and/or optical phase variations. 
After masking out the transits, the BLS algorithm was used in order to search for additional transiting planets, but no additional 
signals were found. Given the lightcurve precision is {$138$ ppm (Table \ref{tab:planet-params}), our data rule out any companion 
larger than $\sim 2R_\Earth$ at 3-sigma with periods $P\lesssim 39$ days}. {As for secondary eclipses, we ran an eclipse fit 
at the expected times, fixing all parameters except for the planet-to-star flux ratio, $F_p/F_*$ and a time shift from the expected 
eclipse times from our circular model in order to allow departures from non-circularity present in the secondary eclipses (and not detected on our 
RV analysis). The result of our fit is presented in Figure \ref{fig:sec_plot}. The retrieved flux ratio in our fit was $F_p/F_* = 35.7^{+16.8}_{-19.0}$ 
ppm and the time shift from the expected secondary eclipse with our circular model (i.e., $T_0-P/2$) was $-1.74^{+0.52}_{-0.24}$ hours, which is 
consistent with a relatively weak detection of a secondary eclipse. This is interesting, however, as it allows us to put a constraint on 
the geometric albedo of the planet. Following \cite{HD2013}, we estimate the geometric albedo of the planet as
\begin{eqnarray*}
A_g = \left[\frac{F_p}{F_*} - \frac{\pi \int_{\lambda_1}^{\lambda_2} B_\lambda (T_{\textnormal{eq}})d\lambda}{F}\left(\frac{R_p}{R_*}\right)\right]\left(\frac{a}{R_p}\right)^2,
\end{eqnarray*}
where $F$ is the irradiation level at the substellar point calculated in Table \ref{tab:planet-params} for our planet, 
$B_\lambda (T_{\textnormal{eq}})$ is a blackbody of temperature $T_{\textnormal{eq}}$ (i.e., we approximate the thermal emission 
of the planet by a blackbody), with
\begin{eqnarray*}
T_{\textnormal{eq}} = T_* \left(\frac{R_* f}{a}\right)^{1/2}(1-A_B)^{1/4},
\end{eqnarray*}
where $f$ is the efficiency of heat redistribution from the dayside to the nightside of the planet and $A_B$ is the Bond albedo. We assume a 
Lambertian sphere (i.e., isotropic scattering), so $A_B = 3A_g/2$ in our calculations. With this, we consider contributions from both 
reflected light (first term in the equation for $A_g$) and thermal emission (second term) from the planet. Integrating the blackbody from 
$\lambda_1 = 0.4\mu$m to $\lambda_2 = 0.9\mu$m (i.e., over the Kepler bandpass), we constrain the geometric albedo to be 
$A_g = 0.47^{+0.12}_{-0.16}$ if we assume complete heat redistribution (i.e., $f = 1/2$), and $A_g = 0.16^{+0.05}_{-0.06}$ if no 
redistribution is assumed (i.e., $f = 2/3$). These values are within the expected geometric albedos of other giant planets measured by the 
Kepler spacecraft \citep[see, e.g.,][]{HD2013,ANGER2015}. Fixing the time shift to zero gives the same constraint on the geometric albedo. 
Optical phase variations were not detected in the lightcurve.}

\section{Discussion}
\label{sec:discussion}

\begin{figure*}
   \includegraphics[width=\columnwidth]{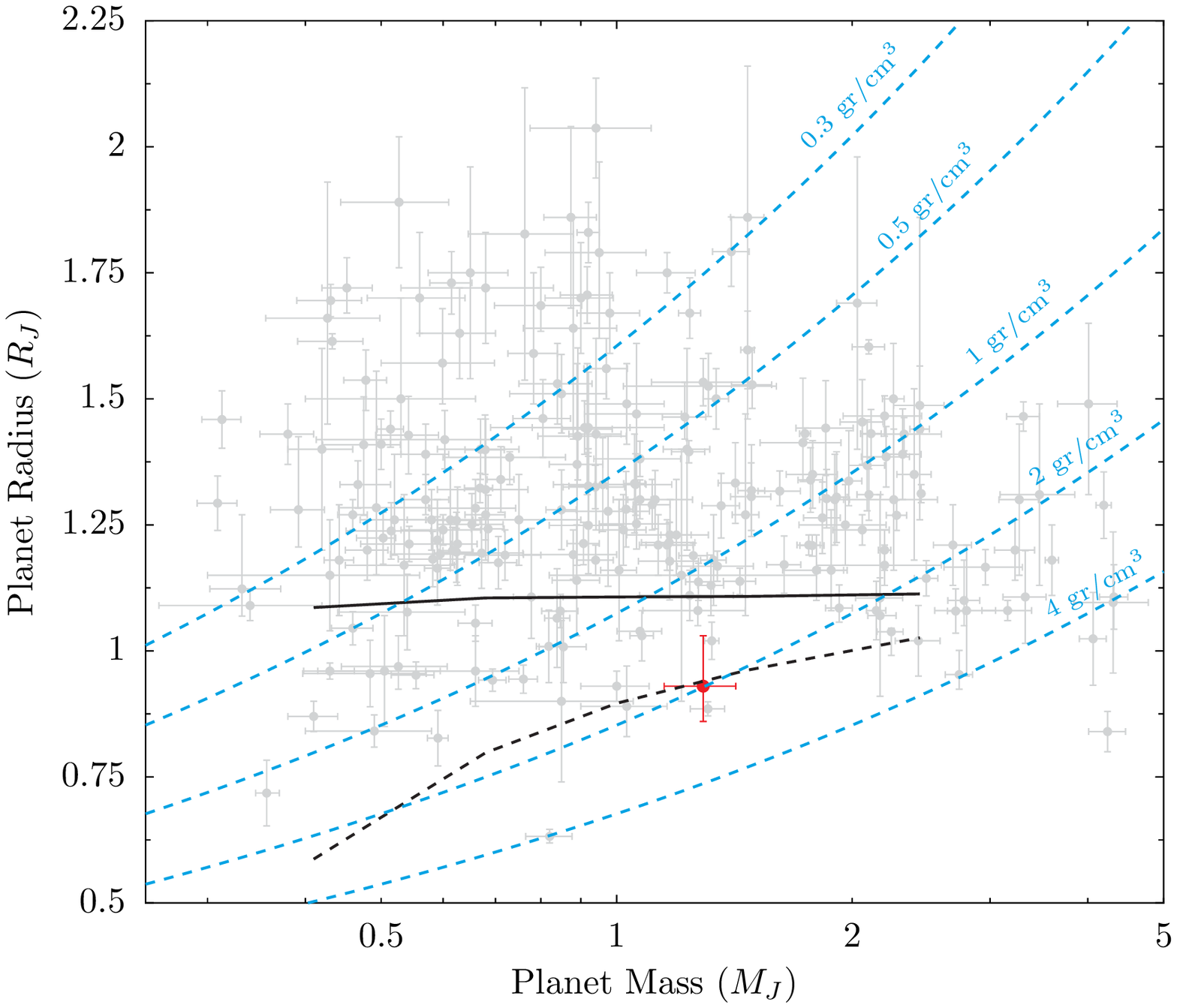}
   \includegraphics[width=\columnwidth]{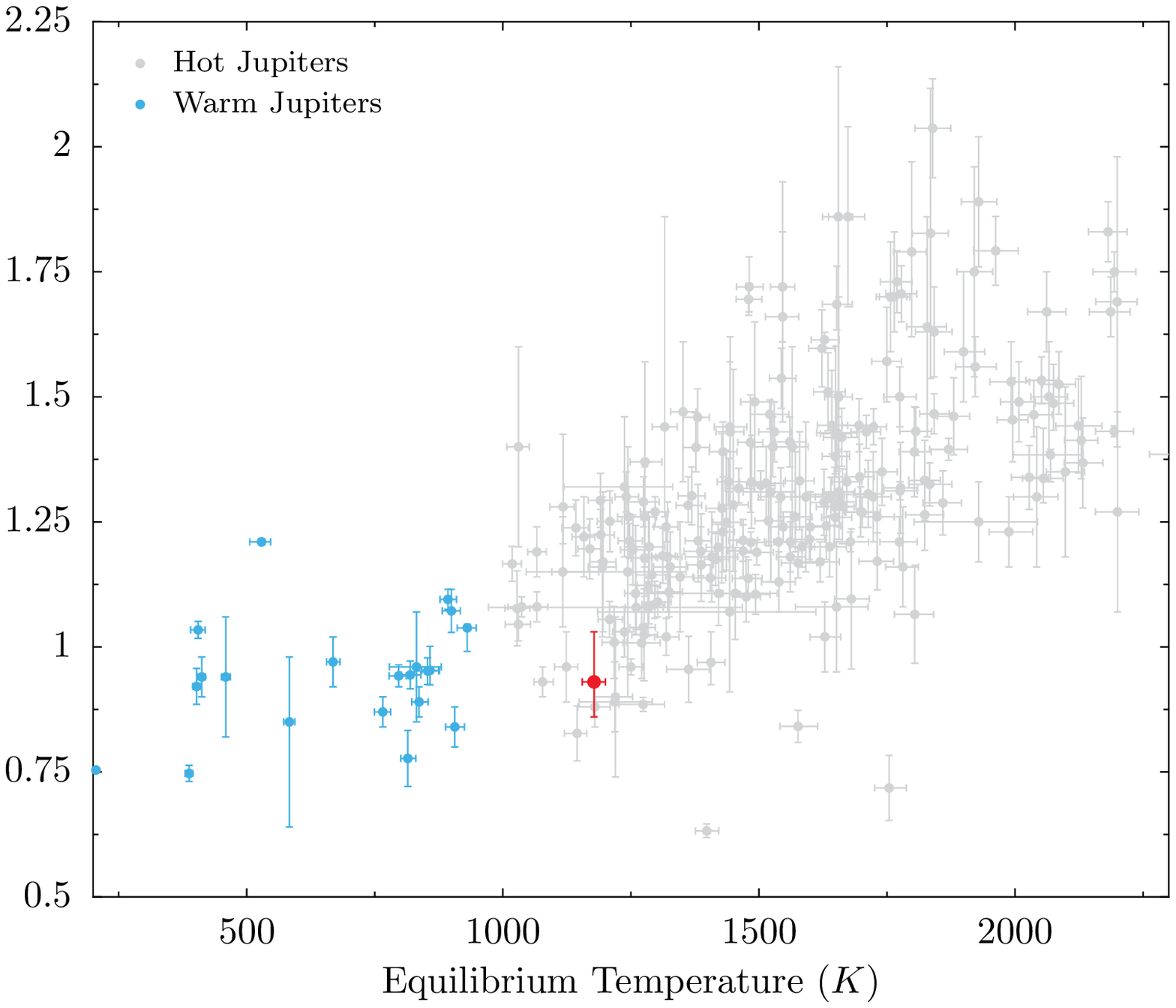}
    \caption{\textit{Left.} Mass-radius diagram for known hot-Jupiters (grey points), along with iso-density curves (blue dashed lines) and the expected 
mass-radius relation for 4.5 Gyr planets at $0.045$ AU from the Sun with no core (black, dashed line) and a $100M\Earth$ core (black, solid line). K2-113b is indicated as the red point with errorbars. \textit{Right.} Equilibrium temperature-radius diagram of known hot-Jupiters (grey points) and 
warm Jupiters (blue points), for comparison. Again, K2-113b is depicted as a red point with errorbars.}
    \label{fig:mrd}
\end{figure*}

Figure \ref{fig:mrd} puts the newly discovered planet in the context of the population of known 
hot-Jupiters in the mass-radius diagram (left panel, $P\lesssim 10$ days, $M\gtrsim 0.1 M_J$) and of the known hot and warm Jupiters in the equilibrium 
temperature-radius diagram (right panel). As can be observed, K2-113b falls on a region in the mass-radius diagram that is currently not very well populated, and which hosts the densest hot-Jupiters with masses below $2M_J$ ($\sim 2$ gr/cm$^3$). In the equilibrium 
temperature-radius diagram, on the other hand, it falls on the typical sizes of warm Jupiters, despite the fact that K2-113b would be 
typically classified as being ``hot" due to its orbit-averaged flux of $\langle F \rangle = 4 \times 10^8$ ergs/cm$^2$/s, which is above (but 
very close to) the $2\times 10^8$ ergs/cm$^2$/s threshold where it is believed ``inflation" mechanisms of giant planets stop being important 
\citep{miller:2011,demory:2011}. Using the relations of \cite{enoch:2012}, K2-113b would be expected to have a radius of $\sim 1.11R_J$, which 
is anyways consistent (at $2$-sigma) with the measured radius. 

The mass and radius of K2-113b could be explained in terms of the ammount of heavy elements in the planet. In Figure \ref{fig:mrd}, we see 
that our planet falls just where the planet evolution models of \cite{fortney:2007} predict it to be if it had a $100 M_\Earth$ core, which is a proxy 
for the ammount of heavy elements in the planet. Of course, giant planets are probably not just H/He envelopes sitting on top of a heavy-element core. 
As shown by \cite{thorngren:2016}, heavy-element enrichment of the envelope is also a very important factor to take into account, although difficult to 
estimate based on planetary mass alone due to the high scatter in the planetary mass-heavy element relation derived in that work{, which 
is both due to the errors on the masses, radii and ages of the planets used to derive that relation, and the stochasticity of the planet formation 
process, which allow for planets of similar mass to have inherently different heavy element content. For example, using this relation, where a 
$10 M_\Earth$ heavy-element core is assumed, the ammount of heavy elements present in the envelope of K2-113b could be anywhere from 
$\sim30 M_\Earth$ to $\sim120 M_\Earth$ at 1-sigma.}

Instead of trying to estimate the heavy element mass in K2-113b directly {which is rather hard to do due to the possibility that 
inflation could be impacting on K2-113b's radius}, we can compare it in terms of its radius and mass to the ``warm" Jupiter WASP-130b 
\citep[$M_p = 1.23\pm 0.04M_J$ and $R_p =0.89\pm 0.03R_J$][]{hellier:2016}. This comparison is interesting because WASP-130b is probably not affected 
by any inflation mechanisms due to its {relatively} low irradiation level {($\log \langle F \rangle = 10^8$, with $F$ in 
cgs units)}. Because of this, \cite{thorngren:2016} was able to use structure models in order to estimate a heavy element mass of $\sim 110M_\Earth$ 
for WASP-130b based on its mass{, radius and age}. Assuming that the ages of both systems are similar, we can use the heavy element 
mass estimated for WASP-130b as a lower limit on the heavy element mass of K2-113b{. This is because, as discussed above, 
K2-113b is within the regime where inflation mechanisms are expected to act, and, thus, the observed radius of K2-113b should be 
larger to what contraction models as the ones used by \cite{thorngren:2016} would predict for its heavy element mass, total mass and age. Following 
this logic, if K2-113b's radius without this inflation mechanism were, say, the radius of WASP-130b, then with inflation the observed radius 
of K2-113b should be larger than WASP-130b's. This is why the estimated heavy element content on WASP-130b is a lower limit on the heavy 
element content of K2-113b -- of course assuming the observed mass, radius and age of both planets are indistinguishable; a reasonable assumption 
given the error bars on those properties.}

{Following a similar logic to the one used above, we can also estimate an upper limit on the heavy element content of K2-113b by comparing it 
to} CoRoT-13b \citep[$M_p = 1.31\pm 0.07$ and $R_p = 0.885\pm 0.014$][]{cabrera:2010}, which is the closest planet in mass and radius to K2-113b among the known hot-Jupiters, despite the fact that the former orbits a hotter star and, hence, has a larger equilibrium temperature 
(1700 K). This difference again could be explained {(if we assume again the system ages, masses and radii are indistinguishable)} 
in terms of the ammount of heavy elements in these planets, with K2-113b having a lower ammount than CoRoT-13b, which is estimated to have 
between $\sim 140-300M_\Earth$ of heavy elements.

The difference in heavy element content between 
WASP-130b, K2-113b and CoRoT-13b would most likely be a signature of their different formation histories rather than a correlation with other 
physical parameters of the system, such as the metallicities of the parent stars, whose correlation with the heavy element content on a given 
planet is rather weak \citep{thorngren:2016}. In fact, in this case the metallicity of WASP-130 is the largest of the three, while the metallicity of 
CoRoT-13 is the smallest, which casts further doubts on the prediction power of such a correlation if it were to exist. 

\section{Conclusions}
\label{sec:conclusions}
In this work, we have presented the discovery of K2-113b, a new hot-Jupiter orbiting a slightly metal rich solar analogue discovered
using photometry from Campaign 8 of the K2 mission and follow-up radial velocities using the FEROS spectrograph. The planet has a radius of
$R_p=0.91^{+0.10}_{-0.07}R_\textnormal{J}$, and a mass of $M_p = 1.28^{+0.11}_{-0.12}M_\textnormal{J}$. With a density of $2.08^{+0.66}_{-0.57}$ 
gr/cm$^3$, the planet is denser than most hot-Jupiters with masses under $2M_J$. We explain its mass and radius in terms of the amount of heavy 
elements in the planet, which should be on the order of $\sim 110 M_\Earth$ or greater. 


\section*{Acknowledgements}
We would like to thank R. Luger for sharing the EVEREST lightcurves for the exoplanet presented in this work as soon as they were available, and 
an anonymous referee for her/his comments which greatly improved the presented work. N.E. is supported by the CONICYT-PCHA/Doctorado Nacional 
graduate fellowship. N.E., R.B., A.J. and C.C. acknowledge support from the Ministry for the Economy, Development, 
and Tourism Programa Iniciativa Cient\'ifica Milenio through grant IC 120009, awarded to the Millennium Institute of Astrophysics (MAS). 
A.J.\ acknowledges support from FONDECYT project 1171208 and from BASAL CATA PFB-06. Support for C. C. is provided by Proyecto FONDECYT 
Iniciaci\'on a la Investigaci\'on 11150768.




\bibliographystyle{mnras}
\bibliography{paperbib} 




%
%


\bsp    
\label{lastpage}
\end{document}